\documentclass[preprintnumbers]{revtex4}
\usepackage{amssymb}
\usepackage{amsmath}
\usepackage{graphicx}
\usepackage{dcolumn}
\usepackage{bm}
\usepackage{subfigure}
\usepackage{color}
\usepackage{CJKutf8}
\numberwithin{equation}{section}
\numberwithin{figure}{section}

\begin{document}
\title{Universality on thermodynamic relation with corrections in higher-dimensional de Sitter black holes}
\author{Hai-Long Zhen,$^{1,2}$ Yu-Bo Ma$^{1,2}$}
\thanks{\emph{e-mail: yuboma.phy@gmail.com}(corresponding author)}

\address{$^1$Department of Physics, Shanxi Datong University, Datong 037009, China\\
$^2$Institute of Theoretical Physics, Shanxi Datong University, Datong, 037009, China\\}

\begin{abstract}
In this study, the methodology proposed by Goon and Penco for investigating the universality on thermodynamic relations with corrections in de Sitter black holes is extended. A universal thermodynamic extremality relation, under consideration of the mass of the spacetime $M$ with various state parameters, proposed by Goon and Penco is investigated in higher dimensional spacetime, the established universal conclusions are not impacted by the convergence of energy from the coexistence region of two horizons to the point $N$ or $C$. Furthermore, by incorporating the shift of the angular momentum into our analysis, a more universal relation is derived, specifically applicable to rotating configurations. Notably, a novel conjecture is formulated that establishes a universal relationship framework connecting shifted thermodynamic quantities across arbitrary black hole backgrounds. These findings are expected to offer profound insights into the fundamental principles of quantum gravity.
\end{abstract}

\maketitle

\section{Introduction}\label{one}

Black holes are of significant consequence in the evolution of the universe. They are capable of devoring matter in their vicinity, releasing considerable quantities of energy in the process. Furthermore, black holes are capable of releasing energy in the form of jets, which can exert an influence on the formation and evolution of galaxies. On 10 April 2019, astronomers worldwide simultaneously achieved a significant milestone in their field by capturing the first-ever photograph of a black hole at the core of the neighbouring galaxy M87, located 55 million light-years from Earth and possessing a mass 6.5 billion times greater than that of the Sun. These supermassive black holes, characterised by their immense size and gravitational pull, represent some of the most enigmatic and destructive objects in the universe \cite{M871,M872}. The discovery of these objects has led to a re-evaluation of the conventional understanding of black hole formation, prompting scientists to reconsider the nature and evolution of black holes. As early as the 1970s, physicists such as Hawking and Bekenstein formulated the four laws of thermodynamics for black holes \cite{HSW,BJD}. The study of the thermodynamic properties of black holes not only contributes to a deeper understanding of the nature of black holes, but also to the understanding of the microscopic states inside them. Despite the extensive research conducted by scientists on black holes, numerous questions pertaining to these enigmatic objects remain unanswered. To illustrate, the information paradox of black holes remains unresolved.

The microstructure of black holes represents a significant theoretical issue that has attracted considerable interest from the scientific community. Despite significant progress in exploring the microstructure of black holes through various methods \cite{044018,111302,071103,101379,2300043,086004,225007,195011,064036}, the internal microstructure of black holes remains ambiguous. To comprehend the quantum properties and microstructure of black holes, it is imperative to first obtain a comprehensive understanding of the thermodynamic properties of black holes. Consequently, the discourse surrounding the diverse thermodynamic characteristics of black holes remains a pivotal subject within the contemporary realm of theoretical physics.

In the field of quantum gravity, the Weak Gravity Conjecture (WGC) as a potential explanatory framework to elucidate the charge-to-mass ratio was proposed by Vafa \cite{V0509212,0601001}. Building on previous studies of the WGC, Goon and Penco investigated the universality of the thermodynamic relation between entropy and extremality under perturbation \cite{101103}. Perturbations in free energy yield a relationship between mass, temperature and entropy, with corrections \cite{00772}. The leading-order expansion of perturbative parameters reveals an approximate relation that can be linked to higher-derivative corrections, thereby establishing a connection between shifts in entropy and the charge-to-mass ratio \cite{046003,0606100,08546}. In light of the foundational relation proposed by Goon and Penco, considerable advancements have been made, encompassing analyses of the relation between diverse AdS spacetimes, including charged BTZ black holes and Kerr-AdS black holes, from the the WGC perspective \cite{14324,08530,07079,08527,115279,11535,00015,L081901,169168}.

It is well established that, for dS spacetime, a coexistence region emerges in the spacetime, encompassing the black hole and cosmological horizons, provided that specific conditions imposed on the parameters in the spacetime are met \cite{Nam418}. The behavior of the mass $M$ of spacetime with respect to $r$ is characterised by two extreme points, corresponding to the masses $M_A$ and $M_C$, respectively. In Ref \cite{17014}, the relationship $M{\rightarrow}M_N$ in the Reissner-Nordstr$\ddot{o}$m-de Sitter (RNdS), Kerr-de Sitter (KdS), and Kerr-Newman-de Sitter (KNdS) spacetime was investigated by employing the theory of Ref \cite{101103}, the universality of the thermodynamic relation was verified in dS black holes. Furthermore, the results suggest that the WGC remains applicable in the dS spacetime.

In the context of the two-horizon coexistence region in dS spacetime, which contains two extremal points, what is the situation regarding the other extremal point? Furthermore, the extremal point $N$ represents the intersection where the positions of the black hole and the cosmological horizon coincide. It can be analyzed along different paths via the black hole horizon or the cosmological horizon when $M{\rightarrow}M_N$ of the two-horizon coexistence region in spacetime. The scenario approaching the extremal point only along the black hole horizon within the two-horizon coexistence region, in case of four-dimensional spacetime case, was addressed in Ref \cite{17014}. To date, no studies have been reported on the approach toward the extremal point along the cosmological horizon or in higher-dimensional spacetimes.
Subsequently, the case of various higher dimensional charged spacetime are analysed in Refs
\cite{139149,4920254,025305,0509212,025108,116581}.
At the moment, it is being proved that Goon and Penco established the relation between the derivative of mass and entropy
\begin{align}\label{1.1}
{{\left( \frac{\partial {{M}_{ext}}(Q,J,\eta )}{\partial \eta } \right)}_{T,Q,J}}=\underset{M\to {{M}_{ext}}}{\mathop{\lim }}\,-T{{\left( \frac{\partial S(M,Q,\eta) }{\partial \eta } \right)}_{M}},
\end{align}
the charge $Q$ and the angular momentum $J$ are considered constants. However, In the presence of a perturbation $\eta$ in spacetime, both $Q$ and $J$ are functions of the perturbation parameter $\eta$ in the higher-dimensional spacetime, in conjunction with a perturbation parameter.

In this study, the universal relation proposed by Goon and Penco in dS black holes is investigated, based on the first law of thermodynamics in dS spacetime \cite{101103}. The methodology of Ref. \cite{17014} is extended to two scenarios: the convergence of extremal points of the two-horizon coexistence region along the black hole horizon and cosmological horizon, respectively; and the convergence to the least extreme point of the two-horizon coexistence region along the black hole horizon. The universal relation that both $Q$ and $J$ are functions of perturbation $\eta$ is given. The study is established on a high-dimensional spacetime with a rotational parameter incorporating a perturbation, the conclusions obtained show a wider theoretical applicability.

The paper is arranged as follows: in Sec. \ref{two}, the universal relation of higher dimensional topological dS black holes with nonlinear source is discussed, and the effects of spacetime dimensions and nonlinear factors on the two-horizon coexistence region are analysed. In Sec. \ref{three}, the universal relation of the $d$-dimensional Kerr-dS spacetimes is investigated by selecting different functions of the spacetime mass $M$ in a universal thermodynamic extremality relation proposed by Goon and Penco. In Sec. \ref{four}, A detailed discussion was conducted on the topic of charged-dS rotating black holes in $n=5$. a more universally applicable Goon-Penco relation is established as a consequence of the consideration of distinct state parameters in the spacetime energy $M$. A brief summary is presented in Sec. \ref{five}

\section{Higher dimensional topological dS black holes with nonlinear source} \label{two}

The $(n+1)$ dimensional action of Einstein gravity with a nonlinear source had been demonstrated in Refs \cite{04863,10309,1950254}, as follows
\begin{align}\label{2.1}
{{I}_{G}}&=-\frac{1}{16\pi }\int\limits_{M}{{{d}^{n+1}}}x\sqrt{-g}[R-2(1+\eta )\Lambda +L(\mathcal{F})]-\frac{1}{8\pi }\int\limits_{\partial M}{{{d}^{n}}}x\sqrt{-\gamma }\Theta (\gamma ), \notag \\
&~~~~~~~~~~~~~~~~~~~~~~~~~~~L(\mathcal{F})=-{\mathcal{F}}+\alpha {{\mathcal{F}}^{2}}+{\mathcal{O}}({{\alpha }^{2}}),
\end{align}
where $R$ and $\Lambda$ are the scalar curvature and the cosmological constant, respectively. $L(\mathcal{F})$ is the Lagrangian of non-linear source with the Maxwell invariant ${\mathcal{F}}=F^{{\mu}{\nu}}F_{{\mu}{\nu}}$, in which $F_{{\mu}{\nu}}=\partial_{\mu}A_{\nu}-\partial_{\nu}A_{\mu}$ is the electromagnetic field tensor and $A_{\mu}$ is the gauge potential. The non-linear charge parameter $\alpha$ is small, so the effects of the non-linear should be regarded as a perturbation. The $(n+1)$-dimensional topological black hole solution is given as follows
\begin{align}\label{2.2}
g(r)=k-\frac{m}{{{r}^{n-2}}}+\frac{2{{q}^{2}}}{(n-1)(n-2){{r}^{2n-4}}}-\frac{(1+\eta ){{r}^{2}}}{{{l}^{2}}}-\frac{4{{q}^{4}}\alpha }{(3{{n}^{2}}-7n+4){{r}^{4n-6}}},
\end{align}
$M$ and $Q$ are the black hole mass and charge, respectively. The curvature radius of dS spacetime is denoted by $l$. The relation $g(r_{+,c})=0$ is satisfied by the horizon position of the black hole, represented by $r_+$, and the cosmological horizon position, represented by $r_c$. The first law of the thermodynamics can be established on the black hole event and cosmological horizons, respectively, as follows
\begin{align}\label{2.3}
dM&={{T}_{+}}d{{S}_{+}}+{{\Phi }_{+}}dQ+{{V}_{+}}dP,\notag \\
dM&=-{{T}_{c}}d{{S}_{c}}+{{\Phi }_{c}}dQ+{{V}_{c}}dP
\end{align}
with
\begin{align}\label{2.4}
M&=\frac{{{V}_{n-1}}(n-1)}{16\pi }\left[ kr_{+,c}^{n-2}-\frac{(1+\eta )r_{+,c}^{n}}{{{l}^{2}}}+\frac{2{{q}^{2}}}{(n-1)(n-2)r_{+,c}^{n-2}} \right.\left. -\frac{4{{q}^{4}}\alpha }{(3{{n}^{2}}-7n+4)r_{+,c}^{3n-4}} \right], \notag \\
{{T}_{+,c}}&=\frac{g'({{r}_{+,c}})}{4\pi }=\frac{\pm }{2\pi (n-1)}\left( \frac{(n-1)(n-2)k}{2{{r}_{+,c}}}-\frac{n(n-1)(1+\eta ){{r}_{+,c}}}{2{{l}^{2}}}-\frac{{{q}^{2}}}{r_{+,c}^{2n-3}}+\frac{2{{q}^{4}}\alpha }{r_{+,c}^{4n-5}} \right), \notag \\
{{S}_{+,c}}&=\frac{{{V}_{n-1}}r_{+,c}^{n-1}}{4},
~~M=m\frac{{{V}_{n-1}}(n-1)}{16\pi },
~~Q=q\frac{{{V}_{n-1}}}{4\pi },
~~{{V}_{n-1}}=\frac{2{{\pi }^{n/2}}}{\Gamma (n/2)}.
\end{align}

\begin{figure}
  \centering
  \subfigure[]{\includegraphics[width=8.8cm,height=4.5cm]{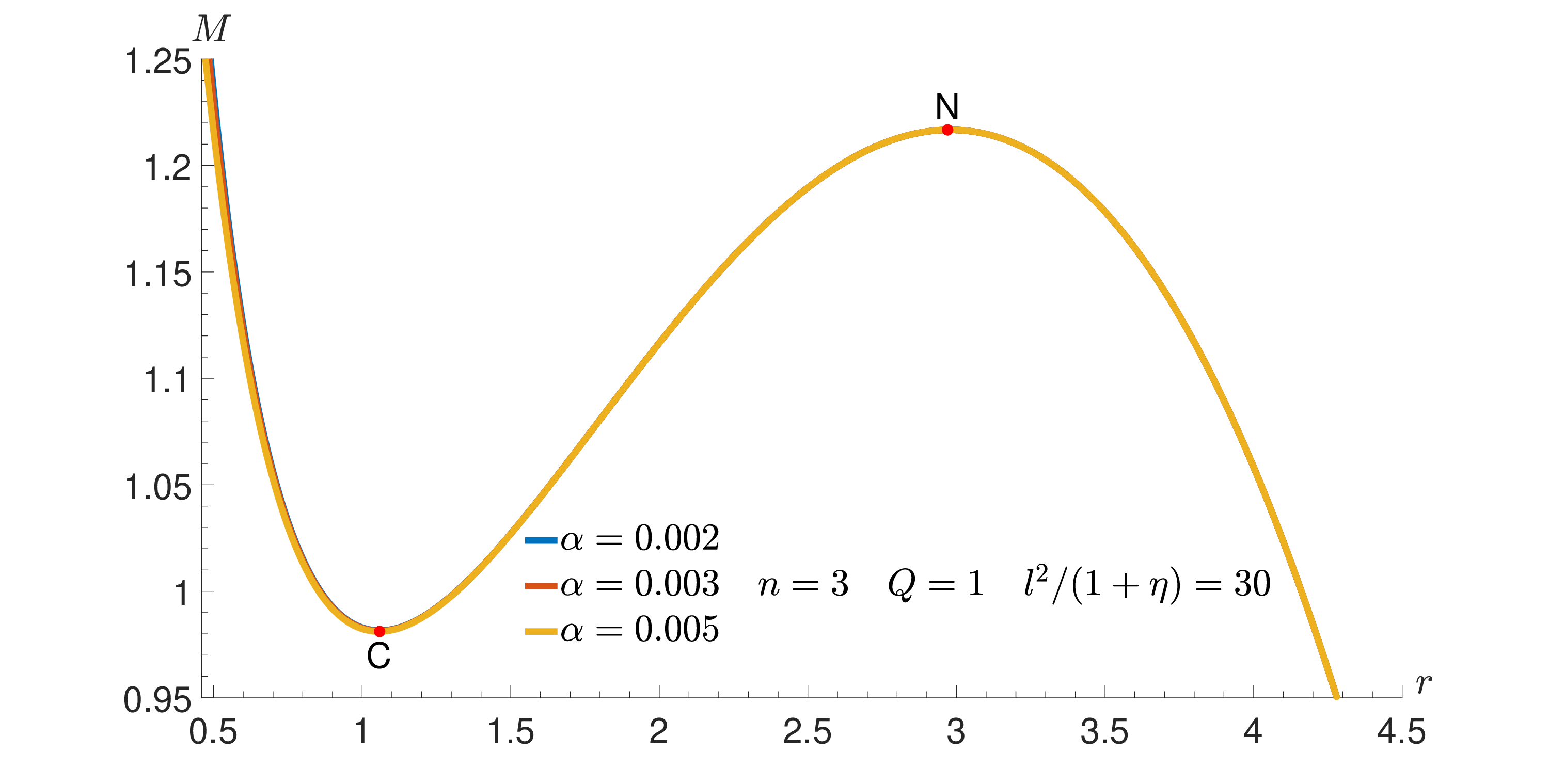}}
  \hspace{0.001cm}%横列子图之间距离
  \centering
  \subfigure[]{\includegraphics[width=8.8cm,height=4.5cm]{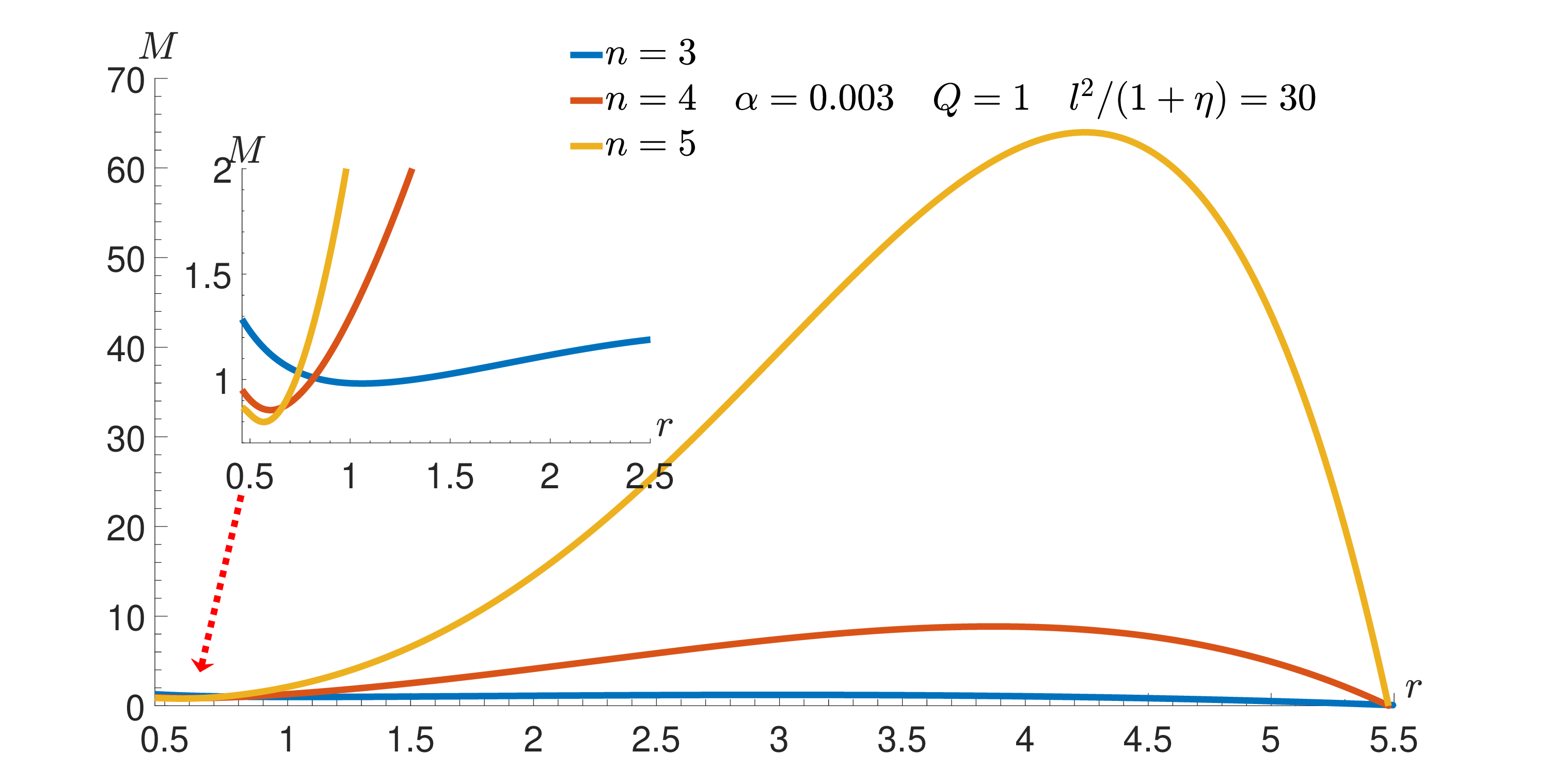}}
  \caption{The $M$-$r$ curve for different $\alpha$ and $n$}\label{Fig2.1}
\end{figure}

As illustrated in Fig. \ref{Fig2.1}(a), it is clear that the effect of the perturbation parameter, $\alpha$, on an inner black hole horizon $r_-$, a black hole horizon $r_+$, and a cosmological horizon $r_c$ in case of the parameters $n=3$, $Q=1$, $l^2/(1+\eta)=30$ is negligible, and the effect on the local maxima $M_N$ and minima $M_C$ is minimal. As demonstrated in Fig. \ref{Fig2.1}(b), the effect of the spacetime dimension $n$ on the inner black hole horizon $r_-$, the black hole horizon $r_+$, and the cosmological horizon $r_c$ is significant for the parameters $\alpha=0.003$, $Q=1$ and $l^2/(1+\eta)=30$. Furthermore,it is evident that the highest point on the curve $M-r$ increases with increasing spacetime dimension $n$. Conversely, the lowest point on the curve $M-r$ decreases as spacetime dimension $n$ increases. This phenomenon suggests that the two-horizon coexistence region undergoes an increase in size with increasing $n$.

The following expression is satisfied by the horizon position of the point $N$ and $C$ in Fig. \ref{Fig2.1}, as follow
\begin{align}\label{2.5}
k(n-2)r_{N,C}^{n-3}-\frac{(1+\eta )nr_{N,C}^{n-1}}{{{l}^{2}}}-\frac{2{{q}^{2}}}{(n-1)(n-2)r_{N,C}^{n-1}}+\frac{4{{q}^{4}}\alpha }{(n-1)r_{N,C}^{3n-3}}=0.
\end{align}

When spacetime dimension $n$, the charge $Q$, the curvature radius $l$ and non-linear parameter $\alpha$ are held constant, the alteration in the energy at point $M_{NC}({{M}_{C}}\le {{M}_{NC}}\le {{M}_{N}})$ in response to a change in $\eta$ is represented by
\begin{align}\label{2.6}
\frac{dM}{d\eta }&=\left( \frac{\partial M}{\partial \eta } \right)+\left( \frac{\partial M}{\partial {{r}_{+,c}}}\frac{\partial {{r}_{+,c}}}{\partial \eta } \right)=-\frac{{{V}_{n-1}}(n-1)}{16\pi }\frac{r_{+,c}^{n}}{{{l}^{2}}} \notag \\
&+\frac{{{V}_{n-1}}(n-1)}{16\pi }\frac{\partial }{\partial {{r}_{+,c}}}\left[ kr_{+,c}^{n-2}-\frac{(1+\eta )r_{+,c}^{n}}{{{l}^{2}}}+\frac{2{{q}^{2}}}{(n-1)(n-2)r_{+,c}^{n-2}} \right.\left. -\frac{4{{q}^{4}}\alpha }{(3{{n}^{2}}-7n+4)r_{+,c}^{3n-4}} \right]\frac{\partial {{r}_{+,c}}}{\partial \eta }.
\end{align}

The variation in entropy in response to a change in $\eta$ is represented by
\begin{align}\label{2.7}
\frac{\partial {{S}_{+,c}}}{\partial \eta }=\frac{(n-1){{V}_{n-1}}r_{+,c}^{n-2}}{4}\frac{\partial {{r}_{+,c}}}{\partial \eta }.
\end{align}

When $M$ is held constant, from Eqs. (\ref{2.3}) and (\ref{2.6}), we obtain
\begin{align}\label{2.8}
-{{T}_{+,c}}\frac{\partial {{S}_{+,c}}}{\partial \eta }&=-{{T}_{+,c}}\frac{\frac{r_{+,c}^{n}}{{{l}^{2}}}\frac{(n-1){{V}_{n-1}}r_{+,c}^{n-2}}{4}}{\frac{\partial }{\partial {{r}_{+,c}}}\left[ kr_{+,c}^{n-2}-\frac{(1+\eta )r_{+,c}^{n}}{{{l}^{2}}}+\frac{2{{q}^{2}}}{(n-1)(n-2)r_{+,c}^{n-2}} \right.\left. -\frac{4{{q}^{4}}\alpha }{(3{{n}^{2}}-7n+4)r_{+,c}^{3n-4}} \right]}
=\mp \frac{r_{+,c}^{n}}{{{l}^{2}}}\frac{(n-1){{V}_{n-1}}}{16\pi }.
\end{align}

From Eq. (\ref{2.6}), we have
\begin{align}\label{2.9}
{{\left( \frac{\partial {{M}_{NC}}}{\partial \eta } \right)}_{Q,\alpha ,l}}&=-\frac{{{V}_{n-1}}(n-1)}{16\pi }\frac{r_{+,c}^{n}}{{{l}^{2}}},
\end{align}
namely, in the coexistence region of the two horizons, when $M{\rightarrow}M_{NC}$ along the black hole horizon position, we have
\begin{align}\label{2.10}
{{\left( \frac{\partial {{M}_{NC}}}{\partial \eta } \right)}_{Q,\alpha ,l}}&=\underset{M\to {{M}_{NC}}}{\mathop{\lim }}\,{{\left( -{{T}_{+}}\frac{\partial {{S}_{+}}}{\partial \eta } \right)}_{M,Q,\alpha ,l}},
\end{align}
namely, in the coexistence region of the two horizons, when $M{\rightarrow}M_{NC}$ along the cosmological horizon position, we have
\begin{align}\label{2.11}
{{\left( \frac{\partial {{M}_{NC}}}{\partial \eta } \right)}_{Q,\alpha ,l}}&=\underset{M\to {{M}_{NC}}}{\mathop{\lim }}\,{{\left( {{T}_{c}}\frac{\partial {{S}_{c}}}{\partial \eta } \right)}_{M,Q,\alpha ,l}}.
\end{align}
The results obtained from Eqs. (\ref{2.10}) and (\ref{2.11}) demonstrate that the relationship remains universal with respect to dimensionality of spacetime. Consequently, it can be concluded that the relation holds in in arbitrary dimensionality of spacetime.

As demonstrated in Refs \cite{101103,17014}, the conclusions presented are merely an convergence to the point $N$ along the position of the black hole event horizon. However, it is important to note that the point $N$ is where the positions of the black hole event horizon and the cosmological horizon are equal. Consequently, in the coexistence region, both horizons have the capacity to approach the point $N$. The present study extends the conclusions of Refs \cite{101103,17014} by deriving the relationship (\ref{2.11}) for approaching the point $N$ along the cosmological horizon in the coexistence region.

It has been demonstrated that the point $c$ is the point of intersection between the black hole event horizon and the black hole inner horizon. Consequently, when the energy $M$ converges to ${{M}_{C}}$ along the black hole event horizon and the cosmological horizon of the coexistence region, i.e., $M\to {{M}_{C}}$, repeating the aforementioned calculations, the following can be obtained
\begin{align}\label{2.12}
{{\left( \frac{\partial {{M}_{C}}}{\partial \eta } \right)}_{Q,\alpha ,l}}=\mp \underset{M\to {{M}_{C}}}{\mathop{\lim }}\,{{\left( {{T}_{+,c}}\frac{\partial {{S}_{+,c}}}{\partial \eta } \right)}_{M,Q,\alpha }}.
\end{align}

It is imperative to acknowledge that the significance of the coexistence region of two horizons in the context of the energy $M\to {{M}_{N,C}}$. In the case of $M\to {{M}_{N}}$, it has been established that the positions of the black hole event horizon and the cosmological horizon coincide, the temperature ${{T}_{+,c}}\to 0$. Conversely, when $M\to {{M}_{C}}$, the positions of the black hole event horizon and the cosmological horizon are not coincident, the temperature ${{T}_{+}}\to 0$, and ${{T}_{c}}\neq 0$. As demonstrated in the aforementioned analysis, it is evident that the established universal conclusions are not impacted by the convergence of energy from the coexistence region of two horizons to the point $N$ or $C$.
\begin{align}\label{2.13}
{{\left( \frac{\partial {{M}_{NC}}}{\partial \eta } \right)}_{Q,\alpha ,l}}=\mp \underset{M\to {{M}_{NC}}}{\mathop{\lim }}\,{{\left( {{T}_{+,c}}\frac{\partial {{S}_{+,c}}}{\partial \eta } \right)}_{M,Q,\alpha ,l}}.
\end{align}

\section{The d-dimensional Kerr-dS spacetimes} \label{three}

It has been demonstrated that the $d$-dimensional Kerr-dS spacetimes \cite{104017,161301} are solutions to the Einstein equations
\begin{align}\label{3.1}
{{R}_{ab}}&=\frac{2(1+{\eta}){\Lambda}}{(d-2)}{{g}_{ab}}.
\end{align}

Thus, we obtained the metric in dS spacetime as follows:
\begin{align}\label{3.2}
d{{s}^{2}}&=-W(1-(1+\eta ){{g}^{2}}{{r}^{2}})d{{t}^{2}}+\frac{2m}{U}{{\left( Wdt-\sum\limits_{i=1}^{N}{\frac{{{a}_{i}}{{\mu }_{i}}d{{\varphi }_{i}}}{{{\Xi }_{i}}}} \right)}^{2}} \notag \\
&+\sum\limits_{i=1}^{N}{\frac{{{r}^{2}}+a_{i}^{2}}{{{\Xi }_{i}}}}(\mu _{i}^{2}d\varphi _{i}^{2}+d\mu _{i}^{2})+\frac{Ud{{r}^{2}}}{X-2m}+\varepsilon {{r}^{2}}d{{\nu }^{2}} \notag \\
&+\frac{(1+\eta ){{g}^{2}}}{W(1-(1+\eta ){{g}^{2}}{{r}^{2}}}{{\left( \sum\limits_{i=1}^{N}{\frac{{{r}^{2}}+a_{i}^{2}}{{{\Xi }_{i}}}{{\mu }_{i}}d{{\mu }_{i}}+\varepsilon {{r}^{2}}\nu d\nu } \right)}^{2}},
\end{align}
where
\begin{align}\label{3.3}
2(1+\eta )\Lambda &=(d-1)(d-2){{g}^{2}} \notag \\
W&=\sum\limits_{i=1}^{N}{\frac{\mu _{i}^{2}}{{{\Xi }_{i}}}}+\varepsilon {{\nu }^{2}},
~~X={{r}^{\varepsilon -2}}(1-(1+\eta ){{r}^{2}}{{g}^{2}}\prod\limits_{i=1}^{N}{({{r}^{2}}+a_{i}^{2})}, \notag \\
U&=\frac{Z}{1-(1+\eta ){{r}^{2}}{{g}^{2}})}\left( 1-\sum\limits_{i=1}^{N}{\frac{a_{i}^{2}}{{{r}^{2}}+a_{i}^{2}}} \right),~~
{{\Xi }_{i}}=1+(1+\eta ){{g}^{2}}a_{i}^{2}.
\end{align}
Here, $N=(d-2)/2$, $[A]$ denotes the integer part of $A$, and we have defined $\varepsilon $ to be $1$ for $d$ even and $0$ for odd. The coordinates ${{\mu }_{i}}$ are not independent, but obey the constraint
\begin{align}\label{3.4}
\sum\limits_{i=1}^{N}{\mu _{i}^{2}}+\varepsilon {{\nu }^{2}}&=1
\end{align}
In even dimensions $d=2N+2$, the thermodynamic quantities are calculated as follows. For the black hole and cosmological horizons, we have
\begin{align}\label{3.5}
{{S}_{+,c}}&=\frac{{{A}_{d-2}}}{4}\prod\limits_{i}{\frac{r_{+,c}^{2}+a_{i}^{2}}{{{\Xi }_{i}}}}=\frac{{{A}_{+,c}}}{4},~~
{{A}_{d-2}}=\frac{2{{\pi }^{(d-2)/2}}}{\Gamma (d-2)/2}, \notag \\
M&=\frac{m{{A}_{d-2}}}{4\pi \prod\nolimits_{j}{{{\Xi }_{j}}}}\sum\limits_{i}{\frac{1}{{{\Xi }_{i}}}},~~
{{J}_{i}}=\frac{m{{a}_{i}}{{A}_{d-2}}}{4\pi {{\Xi }_{i}}\prod\nolimits_{j}{{{\Xi }_{j}}}},~~
\Omega _{+,c}^{i}=\frac{(1-(1+\eta ){{g}^{2}}r_{+,c}^{2}){{a}_{i}}}{r_{+,c}^{2}+a_{i}^{2}}, \notag \\
{{T}_{+,c}}&=\pm \frac{{{r}_{+,c}}(1-(1+\eta ){{g}^{2}}r_{+,c}^{2})}{2\pi }\sum\limits_{i}{\frac{1}{r_{+,c}^{2}+a_{i}^{2}}}\pm \frac{1+(1+\eta ){{g}^{2}}r_{+,c}^{2}}{4\pi {{r}_{+,c}}}, \notag \\
2m&=\frac{1}{{{r}_{+,c}}}(1-(1+\eta )r_{+,c}^{2}{{g}^{2}})\prod\limits_{i}{(r_{+,c}^{2}+a_{i}^{2})},
~~{{\Xi }_{i}}=1+(1+\eta ){{g}^{2}}a_{i}^{2},
\end{align}
In odd dimensions $d=2N+1$, the thermodynamic quantities are calculated as follows. For the black hole and cosmological horizons, we have
\begin{align}\label{3.6}
M&=\frac{m{{A}_{d-2}}}{4\pi \prod\nolimits_{j}{{{\Xi }_{j}}}}\left( \sum\limits_{i}{\frac{1}{{{\Xi }_{i}}}-\frac{1}{2}} \right),\notag \\
{{T}_{+,c}}&=\pm \frac{{{r}_{+,c}}(1-(1+\eta ){{g}^{2}}r_{+,c}^{2})}{2\pi }\sum\limits_{i}{\frac{1}{r_{+,c}^{2}+a_{i}^{2}}}\mp \frac{1}{2\pi {{r}_{+,c}}}, \notag \\
{{S}_{+,c}}&=\frac{{{A}_{d-2}}}{4{{r}_{+,c}}}\prod\limits_{i}{\frac{r_{+,c}^{2}+a_{i}^{2}}{{{\Xi }_{i}}}}=\frac{{{A}_{+,c}}}{4},
\end{align}
The first law of thermodynamics is satisfied by the thermodynamic quantities corresponding to the black hole and cosmological horizons, as follows
\begin{align}\label{3.7}
dM&=\pm {{T}_{+,c}}d{{S}_{+,c}}+\sum\limits_{j}{(\Omega _{+,c}^{j}}-\Omega _{\infty }^{j})d{{J}^{j}}+{{V}_{+,c}}dP
\end{align}
where, the quantities $\Omega _{\infty }^{j}$ allow for the possibility of a rotating frame at infinity \cite{104017,161301,0408217}.
The position $r$ of the highest point $N$ and the lowest point $C$ between the coexistence region of two horizons in spacetime is satisfied by $\frac{\partial M}{\partial r}=0$.
In the event of a perturbation to $\eta $ whilst $g$ is held constant, and when the state parameter $M=M({{S}_{+,c}},J{{(\eta )}_{i(i=1,2,\cdots N)}},\eta )$ is taken into consideration,
${{J}_{i(i=1,2\cdots N)}}(\eta )$ is a function of $\eta $ according to Eq. (\ref{3.5}). Consequently, the universal relationship of the mass $M$ as a function of $\eta $ is given as follows
\begin{align}\label{3.8}
{{\left( \frac{dM}{d\eta } \right)}_{g}}&=\frac{\partial M}{\partial \eta }+\frac{\partial M}{\partial {{S}_{+,c}}}\frac{\partial {{S}_{+,c}}}{\partial \eta }+\sum\limits_{i}{\frac{\partial M}{\partial {{J}_{i}}}\frac{\partial {{J}_{i}}}{\partial \eta }},
\end{align}

According to Eq. (\ref{3.7}), we can obtain
\begin{align}\label{3.9}
{{\left( \frac{dM}{d\eta } \right)}_{g}}&={{\left( \frac{\partial M}{\partial \eta } \right)}_{g,{{S}_{+,c}},J{{(\eta )}_{i(i=1,2,\cdots N)}},M}} \notag \\
&\pm {{T}_{+,c}}{{\left( \frac{\partial {{S}_{+,c}}}{\partial \eta } \right)}_{g,J{{(\eta )}_{i(i=1,2,\cdots N)}},M}}+{{\sum\limits_{j}{(\Omega _{+,c}^{j}-\Omega _{\infty }^{j})\left( \frac{\partial {{J}_{j}}}{\partial \eta } \right)}}_{g,{{S}_{+,c}},J{{(\eta )}_{i\ne j(i=1,2,\cdots N)}},M}}
\end{align}

When $M\to {{M}_{N}}$, $g$ and $J{{(\eta )}_{i(i=1,2,\cdots N)}}$ are held constant according to Eq. (\ref{3.9}), the following equation is obtained
\begin{align}\label{3.10}
{{\left( \frac{\partial {{M}_{NC}}}{\partial \eta } \right)}_{g,J{{(\eta )}_{i(i=1,2,\cdots N)}}}}=\underset{M\to {{M}_{NC}}}{\mathop{\lim }}\,\mp {{T}_{+,c}}{{\left( \frac{\partial {{S}_{+,c}}}{\partial \eta } \right)}_{g,J{{(\eta )}_{i(i=1,2,\cdots N)}},M}},
\end{align}

When $M\to {{M}_{N}}$, ${{S}_{+,c}}$, $g$ and $J{{(\eta )}_{j}}$ are held constant according to Eq. (\ref{3.9}), the following equation is obtained
\begin{align}\label{3.11}
{{\left( \frac{\partial {{M}_{NC}}}{\partial \eta } \right)}_{g,{{S}_{+.c}},J{{(\eta )}_{i\ne j(i=1,2,\cdots N)}}}}=-\underset{M\to {{M}_{NC}}}{\mathop{\lim }}\,(\Omega _{+,c}^{j}-\Omega _{\infty }^{j}){{\left( \frac{\partial {{J}_{j}}}{\partial \eta } \right)}_{g,{{S}_{+,c}},J{{(\eta )}_{i\ne j(i=1,2,\cdots N)}},M}},
\end{align}
in accordance with the Ref \cite{025305}.

When $M\to {{M}_{NC}}$, ${{S}_{+,c}}$ and $g$ are held constant, the following equation is obtained

\begin{align}\label{3.12}
{{\left( \frac{\partial {{M}_{NC}}}{\partial \eta } \right)}_{g,{{S}_{+,c}}}}=-\underset{M\to {{M}_{NC}}}{\mathop{\lim }}\,\sum\limits_{j=1}^{N}{(\Omega _{+,c}^{j}-\Omega _{\infty }^{j}){{\left( \frac{\partial {{J}_{j}}}{\partial \eta } \right)}_{g,{{S}_{+,c}},J{{(\eta )}_{i\ne j(i=1,2,\cdots N)}},M}}}.
\end{align}
It is demonstrated that Eqs. (\ref{3.10}), (\ref{3.11}) and (\ref{3.12}) are satisfied by a universal thermodynamic extremality relation proposed by Goon and Penco when the mass $M$ of the spacetime selected different state parameters are taken into account.

\section{Charged-dS rotating black hole in $n=5$} \label{four}

Finally, the relation will be verified for a charged-dS rotating black hole in $n=5$. The aforementioned solution comprises axially symmetric black holes with an axis of rotation. It is important to note that the action is identical to the SdS action, leading to a process similar to the SdS case:
The $n=5$ Kerr-Newman-de Sitter metric for a rotating charged black hole with a positive cosmological constant is given by \cite{104017,161301}
\begin{align}\label{4.1}
d{{s}^{2}}&=-\frac{R[(1+{{g}^{2}}{{r}^{2}}){{\rho }^{2}}dt+2q\nu ]dt}{{{\Xi }_{a}}\Xi {}_{b}{{\rho }^{2}}}+\frac{2q\nu \omega }{{{\rho }^{2}}}+\frac{f}{{{\rho }^{4}}}{{\left( \frac{Rdt}{{{\Xi }_{a}}\Xi {}_{b}}-\omega  \right)}^{2}}\notag \\
&+\frac{{{\rho }^{2}}d{{r}^{2}}}{\Delta }+\frac{{{\rho }^{2}}d{{\theta }^{2}}}{R}+\frac{{{r}^{2}}+{{a}^{2}}}{{{\Xi }_{a}}}{{\sin }^{2}}\theta d{{\varphi }^{2}}+\frac{{{r}^{2}}+{{b}^{2}}}{{{\Xi }_{b}}}{{\cos }^{2}}\theta d{{\psi }^{2}},\notag \\
\phi &=\frac{\sqrt{3}q}{{{\rho }^{2}}}\left( \frac{Rdt}{{{\Xi }_{a}}{{\Xi }_{b}}}-\omega  \right)
\end{align}
where
\begin{align}\label{4.2}
\nu &=b{{\sin }^{2}}\theta d\varphi +a{{\cos }^{2}}\theta d\psi ,~~\omega =a{{\sin }^{2}}\theta \frac{d\varphi }{{{\Xi }_{a}}}+b{{\cos }^{2}}\theta \frac{d\psi }{{{\Xi }_{b}}},\notag \\
R&=1+{{a}^{2}}{{g}^{2}}{{\cos }^{2}}\theta +{{b}^{2}}{{g}^{2}}{{\sin }^{2}}\theta ,
~~\Delta =\frac{({{r}^{2}}+{{a}^{2}})({{r}^{2}}+{{b}^{2}})(1-{{g}^{2}}{{r}^{2}})+{{q}^{2}}+2abq}{{{r}^{2}}}-2m, \notag \\
{{\rho }^{2}}&=1+b{{\sin }^{2}}\theta +a{{\cos }^{2}}\theta ,
~~{{\Xi }_{a}}=1+{{a}^{2}}{{g}^{2}},~~{{\Xi }_{b}}=1+{{b}^{2}}{{g}^{2}},
~~f=2m{{\rho }^{2}}-{{q}^{2}}-2abq{{g}^{2}}{{\rho }^{2}},
\end{align}
The thermodynamic quantities are \cite{025108}
\begin{align}\label{4.3}
M&=\frac{\pi m(2{{\Xi }_{a}}+2{{\Xi }_{b}}-{{\Xi }_{a}}{{\Xi }_{b}})-2\pi qab{{g}^{2}}({{\Xi }_{a}}+{{\Xi }_{b}})}{4\Xi _{a}^{2}\Xi _{b}^{2}}, \notag \\
m&=\frac{({{r}^{2}}+{{a}^{2}})({{r}^{2}}+{{b}^{2}})(1-{{g}^{2}}{{r}^{2}})+{{q}^{2}}+2abq}{2{{r}^{2}}}, \notag \\
{{J}^{a}}&=\frac{\pi [2am+qb(1-{{a}^{2}}{{g}^{2}})]}{4\Xi _{a}^{2}{{\Xi }_{b}}},
~~{{J}^{b}}=\frac{\pi [2bm+qb(1-{{a}^{2}}{{g}^{2}})]}{4\Xi _{b}^{2}{{\Xi }_{a}}},
~~Q=\frac{\sqrt{3}\pi q}{4{{\Xi }_{a}}{{\Xi }_{b}}},
\end{align}
where $r$ is used to represent the black hole horizon, $r_+$ or the cosmological horizon, $r_c$. For the black hole horizon, we have
\begin{align}\label{4.4}
{{T}_{+}}&=\frac{r_{+}^{4}[1-{{g}^{2}}(2r_{+}^{2}+{{a}^{2}}+{{b}^{2}})]-{{(ab+q)}^{2}}}{2\pi {{r}_{+}}[(r_{+}^{2}+{{a}^{2}})(r_{+}^{2}+{{b}^{2}})+abq]},
~~{{S}_{+}}=\frac{{{\pi }^{2}}[(r_{+}^{2}+{{a}^{2}})(r_{+}^{2}+{{b}^{2}})+abq]}{2{{\Xi }_{a}}{{\Xi }_{b}}{{r}_{+}}}, \notag \\
{{\phi }_{+}}&=\frac{\sqrt{3}qr_{+}^{2}}{(r_{+}^{2}+{{a}^{2}})(r_{+}^{2}+{{b}^{2}})+abq},
~~\Omega _{+}^{a}=\frac{a(r_{+}^{2}+{{b}^{2}})(1-{{g}^{2}}r_{+}^{2})+bq}{(r_{+}^{2}+{{a}^{2}})(r_{+}^{2}+{{b}^{2}})+abq}, \notag \\
\Omega _{+}^{b}&=\frac{b(r_{+}^{2}+{{a}^{2}})(1-{{g}^{2}}r_{+}^{2})+aq}{(r_{+}^{2}+{{a}^{2}})(r_{+}^{2}+{{b}^{2}})+abq},
\end{align}
For the cosmological horizon, we have
\begin{align}\label{4.5}
{{T}_{c}}&=-\frac{r_{c}^{4}[1-{{g}^{2}}(2r_{c}^{2}+{{a}^{2}}+{{b}^{2}})]-{{(ab+q)}^{2}}}{2\pi {{r}_{c}}[(r_{c}^{2}+{{a}^{2}})(r_{c}^{2}+{{b}^{2}})+abq]},
~~{{S}_{c}}=\frac{{{\pi }^{2}}[(r_{c}^{2}+{{a}^{2}})(r_{c}^{2}+{{b}^{2}})+abq]}{2{{\Xi }_{a}}{{\Xi }_{b}}{{r}_{c}}}, \notag \\
{{\phi }_{c}}&=\frac{\sqrt{3}qr_{c}^{2}}{(r_{c}^{2}+{{a}^{2}})(r_{c}^{2}+{{b}^{2}})+abq},
~~\Omega _{c}^{a}=\frac{a(r_{c}^{2}+{{b}^{2}})(1-{{g}^{2}}r_{c}^{2})+bq}{(r_{c}^{2}+{{a}^{2}})(r_{c}^{2}+{{b}^{2}})+abq}, \notag \\
\Omega _{c}^{b}&=\frac{b(r_{c}^{2}+{{a}^{2}})(1-{{g}^{2}}r_{c}^{2})+aq}{(r_{c}^{2}+{{a}^{2}})(r_{c}^{2}+{{b}^{2}})+abq},
\end{align}
the first law of the thermodynamics is established on the black hole event and cosmological horizons, as follows
\begin{align}\label{4.6}
dM=\pm {{T}_{+,c}}d{{S}_{+,c}}+\sum\limits_{j}{\Phi _{+,c}^{j}}d{{Q}^{j}}+\sum\limits_{j}{(\Omega _{+,c}^{j}-\Omega _{\infty }^{j}})d{{J}^{j}}+{{V}_{+,c}}dP,
\end{align}
where the quantities $\Omega _{\infty }^{j}$, allow for the possibility of a rotating frame at infinity \cite{104017,161301,0408217}, the position of the black hole event and cosmological horizon, denoted by $r_{+,c}$. According to \cite{0606100}, the cosmological constant becomes the perturbation parameter $\eta$ being a very small constant \cite{101103,4920254,139149}.

\begin{figure}[htb]
\includegraphics[width=9.0cm,height=4.5cm]{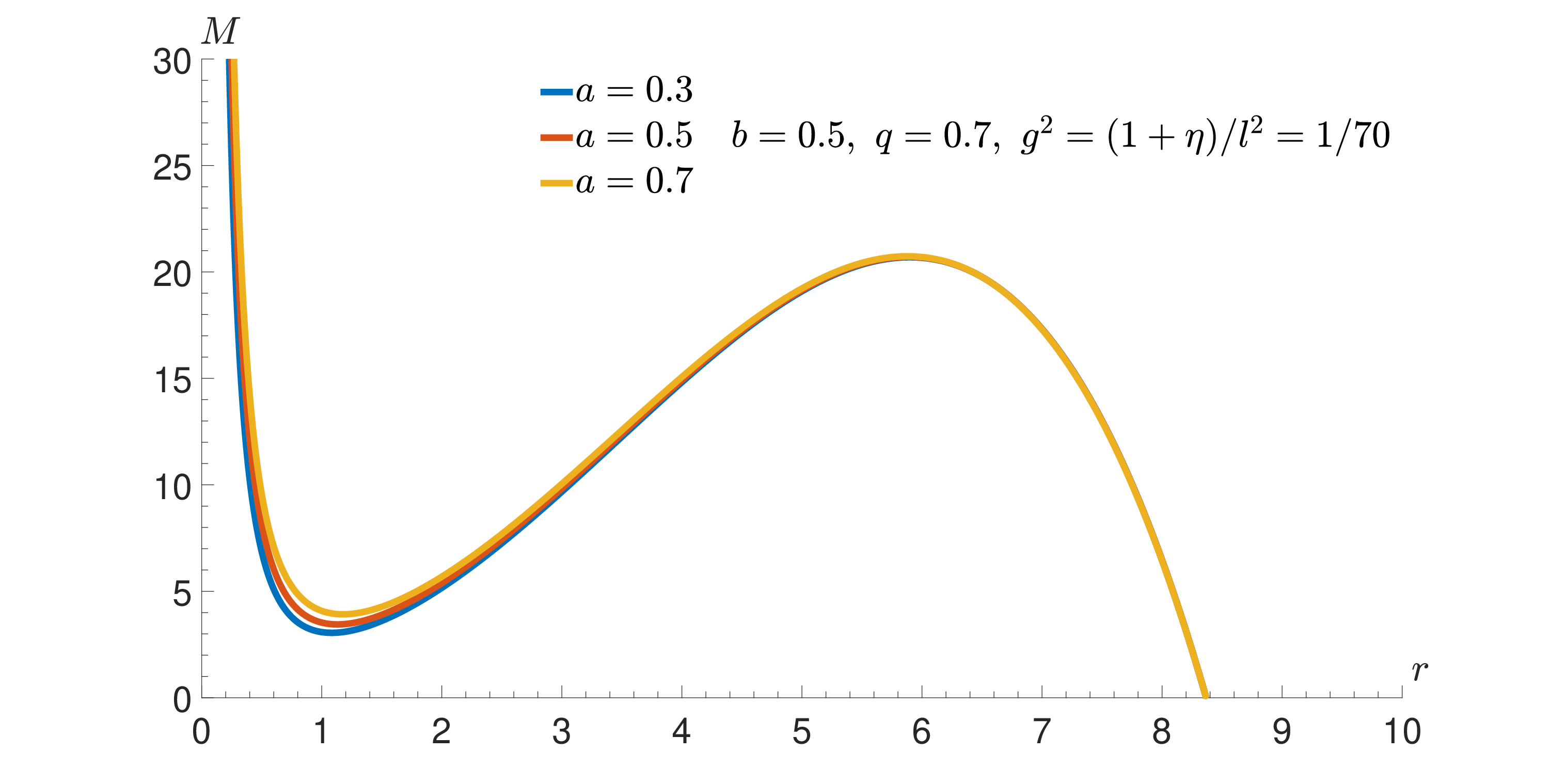}
\vskip -4mm \caption{The $M$-$r$ curve}\label{Fig4.1}
\end{figure}

The behavior between the position of the horizon, denoted by $r_{+,c}$, and the mass, denoted by $M$ can be illustrated by plotting the $M-r$ curve in Fig. \ref{Fig4.1} according to Eq. (\ref{4.3})

As demonstrated in Fig. \ref{Fig4.1}, it is evident from that the actions of $a$ and $b$ are equal. Furthermore, the highest point $N$ of the curve remains unaltered by the values of $a$ and $b$. Conversely, the lowest point $C$ of the curve increases as $a$ and $b$ increase. This phenomenon suggests that the coexistence region of the two horizons undergoes a decrease in size.

From Fig. \ref{Fig4.1}, it can be seen that no horizon exist in spacetime when $M>M_N$ \cite{Nam418}. The position of the black hole horizon, $r_+$, and the cosmological horizon, $r_c$, coincide when $M=M_N$, the position of the horizon is satisfied by
\begin{align}\label{4.7}
\frac{\partial M}{\partial r}&=\frac{\pi (2{{\Xi }_{a}}+2{{\Xi }_{b}}-{{\Xi }_{a}}{{\Xi }_{b}})}{4\Xi _{a}^{2}\Xi _{b}^{2}}\frac{\partial m}{\partial r}=\frac{(2{{r}^{2}}+{{a}^{2}}+{{b}^{2}})(1-{{g}^{2}}{{r}^{2}})-{{g}^{2}}({{r}^{2}}+{{a}^{2}})({{r}^{2}}+{{b}^{2}})}{r} \notag \\
&=\frac{\pi (2{{\Xi }_{a}}+2{{\Xi }_{b}}-{{\Xi }_{a}}{{\Xi }_{b}})}{4\Xi _{a}^{2}\Xi _{b}^{2}}\frac{{{r}^{4}}[1-{{g}^{2}}(2{{r}^{2}}+{{a}^{2}}+{{b}^{2}})]-{{(ab+q)}^{2}}}{{{r}^{3}}}=0
\end{align}

As $M=M_C$ the inner horizon, denoted by $r_-$, and the event horizon, denoted by $r_+$, coincide together, and such a black hole is called the cold black hole whose event horizon radius, denoted by $r_{cold}$, is the remaining positive real solution of Eq. (\ref{4.6}). As $M_C<M<M_N$ the coexistence region with the black hole horizon, $r_+$, and the cosmological horizon, $r_c$, exists in dS spacetime. As $M<M_N$ the cosmological horizon, $r_c$, only exists in dS spacetime.

It can be demonstrated that the energy, $M$, the entropy, ${{S}_{+,c}}$, the charge, $Q$, as well as the angular momentum, ${{J}_{a}}(\eta )$ and ${{J}_{b}}(\eta )$, of the spacetime are all functions of the perturbation parameter, $\eta$, when $g$ is fixed according to Eq. (\ref{4.3}). Taking $M=M({{S}_{+,c}},{{J}_{a}}(\eta ),{{J}_{b}}(\eta ),Q(\eta ))$, the dimensionality of the spacetime, $n$, and $g$ hold constant, the following expression is provided for the alteration of the energy with respect to the alteration of the perturbation parameter, $\eta$, from Eqs. (\ref{4.4}) and (\ref{4.5})
\begin{align}\label{4.8}
{{\left( \frac{dM}{d\eta } \right)}_{g}}={{\left( \frac{\partial M}{\partial \eta } \right)}_{g,{{S}_{+.c}},Q(\eta ),{{J}_{a}}(\eta ),{{J}_{b}}(\eta )}}+\frac{\partial M}{\partial {{S}_{+,c}}}\frac{\partial {{S}_{+.c}}}{\partial \eta }+\frac{\partial M}{\partial {{J}_{a}}}\frac{\partial {{J}_{a}}}{\partial \eta }+\frac{\partial M}{\partial {{J}_{b}}}\frac{\partial {{J}_{b}}}{\partial \eta }+\frac{\partial M}{\partial Q}\frac{\partial Q}{\partial \eta }.
\end{align}

From Eq. (\ref{3.6}), we have
\begin{align}\label{4.9}
{{\left( \frac{dM}{d\eta } \right)}_{g}}&={{\left( \frac{\partial M}{\partial \eta } \right)}_{g,{{S}_{+.c}},Q(\eta ),{{J}_{a}}(\eta ),{{J}_{b}}(\eta )}}\pm {{T}_{+,c}}{{\left( \frac{\partial {{S}_{+.c}}}{\partial \eta } \right)}_{g,{{J}_{a}}(\eta ),{{J}_{b}}(\eta ),Q(\eta ),M}}+(\Omega _{+,c}^{a}-\Omega _{\infty }^{a}){{\left( \frac{\partial {{J}_{a}}}{\partial \eta } \right)}_{g,{{S}_{+,c}},{{J}_{b}}(\eta ),Q(\eta ),M}} \notag \\
&+(\Omega _{+.c}^{b}-\Omega _{\infty }^{b}){{\left( \frac{\partial {{J}_{b}}}{\partial \eta } \right)}_{g,{{S}_{+,c}},{{J}_{a}}(\eta ),Q(\eta ),M}}+{{\phi }_{+,c}}{{\left( \frac{\partial Q}{\partial \eta } \right)}_{g,{{S}_{+.c}},{{J}_{a}}(\eta ),{{J}_{b}}(\eta ),M}}.
\end{align}

When $M\to {{M}_{NC}}$, ${{J}_{a}}(\eta )$ ${{J}_{b}}(\eta )$ and $Q(\eta )$ are held constant according to Eq. (\ref{4.9}), we have
\begin{align}\label{4.10}
{{\left( \frac{\partial {{M}_{NC}}}{\partial \eta } \right)}_{g,Q(\eta ),{{J}_{a}}(\eta ),{{J}_{b}}(\eta )}}=\underset{M\to {{M}_{NC}}}{\mathop{\lim }}\,\mp {{T}_{+,c}}{{\left( \frac{\partial {{S}_{+,c}}}{\partial \eta } \right)}_{g,{{J}_{a}}(\eta ),{{J}_{b}}(\eta ),Q(\eta ),M}}.
\end{align}

When $M\to {{M}_{NC}}$, ${{S}_{+,c}}$, $g$, $Q(\eta)$ and ${{J}_{a}}(\eta)$ are held constant according to Eq. (\ref{4.9}), we have
\begin{align}\label{4.11}
{{\left( \frac{\partial {{M}_{NC}}}{\partial \eta } \right)}_{g,{{S}_{+,c}},Q(\eta ),{{J}_{a}}(\eta )}}=-\underset{M\to {{M}_{NC}}}{\mathop{\lim }}\,(\Omega _{+,c}^{b}-\Omega _{\infty }^{b}){{\left( \frac{\partial {{J}_{b}}}{\partial \eta } \right)}_{g,{{S}_{+,c}},Q(\eta ),{{J}_{a}}(\eta ),M}}.
\end{align}

When $M\to {{M}_{NC}}$, ${{S}_{+,c}}$, $g$ and $Q(\eta)$ are held constant according to Eq. (\ref{4.9}), we have
\begin{align}\label{4.12}
{{\left( \frac{\partial {{M}_{NC}}}{\partial \eta } \right)}_{g,{{S}_{+,c}},Q(\eta )}}
&=-\underset{M\to {{M}_{NC}}}{\mathop{\lim }}\,\left[ (\Omega _{+,c}^{a}-\Omega _{\infty }^{a}){{\left( \frac{\partial {{J}_{a}}}{\partial \eta } \right)}_{g,{{S}_{+,c}},{{J}_{b}}(\eta ),Q(\eta ),M}}+(\Omega _{+.c}^{b}-\Omega _{\infty }^{b}){{\left( \frac{\partial {{J}_{b}}}{\partial \eta } \right)}_{g,{{S}_{+,c}},{{J}_{a}}(\eta ),Q(\eta ),M}} \right].
\end{align}

When $M\to {{M}_{NC}}$, ${{S}_{+,c}}$, $g$, ${{J}_{a}}(\eta)$ and ${{J}_{b}}(\eta)$ are held constant according to Eq. (\ref{4.9}), we have
\begin{align}\label{4.13}
{{\left( \frac{\partial {{M}_{NC}}}{\partial \eta } \right)}_{g,{{S}_{+,c}},{{J}_{a}}(\eta ),{{J}_{b}}(\eta ))}}=-\underset{M\to {{M}_{NC}}}{\mathop{\lim }}\,{{\varphi }_{+,c}}{{\left( \frac{\partial Q}{\partial \eta } \right)}_{{{S}_{+,c}},g,{{J}_{a}}(\eta ),{{J}_{b}}(\eta ),M}}.
\end{align}

When $M\to {{M}_{NC}}$, ${{S}_{+,c}}$, $g$ and ${{J}_{b}}(\eta)$ are held constant according to Eq. (\ref{4.9}), we have
\begin{align}\label{4.14}
{{\left( \frac{\partial {{M}_{NC}}}{\partial \eta } \right)}_{g,{{S}_{+,c}},{{J}_{b}}(\eta )}}=-\underset{M\to {{M}_{NC}}}{\mathop{\lim }}\,\left[ {{\varphi }_{+,c}}{{\left( \frac{\partial Q}{\partial \eta } \right)}_{g,{{S}_{+,c}},{{J}_{a}}(\eta ),{{J}_{b}}(\eta ),M}}+(\Omega _{+,c}^{a}-\Omega _{\infty }^{a}){{\left( \frac{\partial {{J}_{a}}}{\partial \eta } \right)}_{g,{{S}_{+,c}},{{J}_{b}}(\eta ),Q(\eta ),M}} \right].
\end{align}

When $M\to {{M}_{NC}}$, $g$ and ${{J}_{b}}(\eta)$ are held constant according to Eq. (\ref{4.9}), we have
\begin{align}\label{4.15}
{{\left( \frac{\partial {{M}_{NC}}}{\partial \eta } \right)}_{g,{{J}_{b}}(\eta )}}&=\underset{M\to {{M}_{NC}}}{\mathop{\lim }}\,\mp {{T}_{+,c}}{{\left( \frac{\partial {{S}_{+,c}}}{\partial \eta } \right)}_{g,{{J}_{a}}(\eta ),{{J}_{b}}(\eta ),Q(\eta ),M}} \notag \\
&-\underset{M\to {{M}_{NC}}}{\mathop{\lim }}\,\left[ -{{\varphi }_{+,c}}{{\left( \frac{\partial Q}{\partial \eta } \right)}_{g,{{S}_{+,c}},{{J}_{a}}(\eta ),{{J}_{b}}(\eta ),M}}+(\Omega _{+,c}^{a}-\Omega _{\infty }^{a}){{\left( \frac{\partial {{J}_{a}}}{\partial \eta } \right)}_{g,{{S}_{+,c}},{{J}_{b}}(\eta ),Q(\eta ),M}} \right].
\end{align}

It is evident that the universal Goon-Penco relation is obtained as the energy of the spacetime, $M$, is taken different state parameters into account.

\section{Summary}\label{five}

In the preceding analysis, the primary thesis examines how the behavior of the state parameters and the position of the horizon change in response to variations in the spacetime perturbation paratemeter, $\eta$, while other parameters of spacetime remain constant. The findings align with those reported in Ref \cite{101103}. Additionally, the results from Refs. \cite{101103,17014} are extended to two extreme points, $M_N$ and $M_C$, and the convergence from the coexistence region of two horizons to these points along the cosmological horizon is discussed. In the present study, the charge, $Q$, and the angular momentum, $J$, of spacetime are considered as a function of the perturbation parameter, $\eta$. Furthermore, the study is extended to arbitrary spacetime dimensionality. The methodology developed in this study offers a novel approach to the investigation of the physical properties of the coexistence region between two horizons in dS spacetime, characterised by its computational efficiency and transparency.

The region of coexistence exists between the black hole event horizon and the cosmological horizon in dS spacetime. In this region, the energy associated with the black hole event horizon possesses a maximum value, denoted by $M_N$, and a minimum value, denoted by $M_C$. The universal relationship of Eqs. (\ref{2.13}), (\ref{3.10}) and (\ref{4.10}) are satisfied in spacetime when the energy, $M$, in spacetime is satisfied by $M\le {{M}_{NC}}\le {{M}_{N}}$. The exploration of the proportional relationship between corrected mass and entropy provides a pathway to understanding the WGC. The WGC offers intriguing insights into the realm of quantum gravity, and studies such as the one presented in this paper contribute to a deeper comprehension of this fundamental aspect of physics. To conclude, our calculations show the universality of the relations, at least for the black holes studied.

\section*{Acknowledgments}
We would like to thank Prof. Ren Zhao and Meng-Sen Ma for their indispensable discussions and comments. This work was supported by the Natural Science Foundation of China (Grant No. 12375050, Grant No. 11705106, Grant No. 12075143), the Scientific Innovation Foundation of the Higher Education Institutions of Shanxi Province (Grant Nos. 2023L269)

\end{document}